\documentclass[twocolumn,preprintnumbers,amsmath,amssymb]{revtex4-1}
\usepackage{graphicx}
\usepackage{dcolumn}
\usepackage{bm}
\usepackage[usenames,dvipsnames]{xcolor}

\begin{document}
\preprint{ \color{NavyBlue} DRAFT v2.7 - \today}

\title{\Large \color{NavyBlue}  Strings of droplets propelled by coherent waves}

\author{B.Filoux}
\author{M.Hubert}
\author{N.Vandewalle}
\address{GRASP, Institute of Physics B5a, University of Li\`ege, B4000 Li\`ege, Belgium.}

\maketitle


{\bf 
Bouncing walking droplets possess fascinating properties due to their peculiar wave/particle interaction. In order to study such walkers in a 1d system, we considered the case of a few droplets in an annular cavity. We show that, in this geometry, they spontaneously form a string of synchronized bouncing droplets that share a common coherent wave propelling the group at a speed faster than single walkers. The formation of this coherent wave and the collective droplet behaviors are captured by a model, which sheds a new light on droplet/wave interactions. 
}


When a tiny droplet is gently placed along a liquid interface, which is vibrated with an amplitude $A$ and a frequency $f$, it is able to bounce without coalescing when the bath maximum acceleration $\Gamma = {4\pi^2 A f^2 / g}$ is above a threshold $\Gamma_B$ \citep{Couder2005,Terwagne2013,Hubert2015}. In fact, an air layer separates the droplet from the vibrated surface preventing coalescence. Depending on the forcing parameter $\Gamma$, various bouncing modes can be observed, from simple bounces to period doubling and much more complex trajectories \citep{Wind2013,Protiere2006}. By approaching the Faraday instability from below ($\Gamma_B< \Gamma < \Gamma_F$), the droplet starts to bounce once every two periods, and the emitted waves possess the characteristics of the Faraday pattern with a wavelength $\lambda_F$ being fixed by the forcing parameters \citep{Protiere2006}. 
 
Due to the coupling between the droplet and the sum of waves emitted on the liquid surface $\zeta(\vec r,t)$ by the successive previous impacts, bouncing droplets may start to move horizontally along the liquid \cite{Couder2005}. They are called walkers \citep{Protiere2006}. The droplet-wave interaction leads to spectacular physical phenomena at a macroscopic scale : single walkers may exhibit tunneling effect \citep{Eddi2009}, diffraction and interference around apertures \citep{Couder2006}, revolution orbits \citep{Protiere2008} and splitting of energy levels \citep{Eddi2012}. Resonance into cavities were also investigated \citep{Harris2013}. Processes involving two or more droplets were also investigated but the formation of large coherent waves is still a challenge for obtaining macroscopic collective effects \citep{Eddi2009-b}. We will show herein how to create such coherent waves in a confined geometry. 
 
The experimental conditions for the droplet and the bath are detailed in Methods. Identical droplets are created by an automatic generator \citep{Terwagne2013}. The liquid is silicon oil (viscosity $\eta=20$ cSt). The container is shaken vertically by an electromagnetic system with a tunable amplitude $A$ and a fixed frequency $f=70 \, {\rm Hz}$. The dimensionless maximum acceleration $\Gamma$ is adjusted to find the bouncing and walking regimes close to the Faraday instability. The acceleration is always kept at $\Gamma = 0.95 \Gamma_F$ in our experiments. This corresponds to a so-called ``short memory regime" \citep{Labousse2014} since Faraday waves are damped such that the liquid surface $\zeta(\vec r, t)$ keeps the trace of about ${\rm Me} = \Gamma_F/(\Gamma_F-\Gamma) \approx 20$ previous impacts. The wavelength has been estimated to about $\lambda_F \approx 6 \, {\rm mm}$. 

\begin{figure}[h]
\begin{center}
\vskip 0.1 cm
\includegraphics[width=8cm]{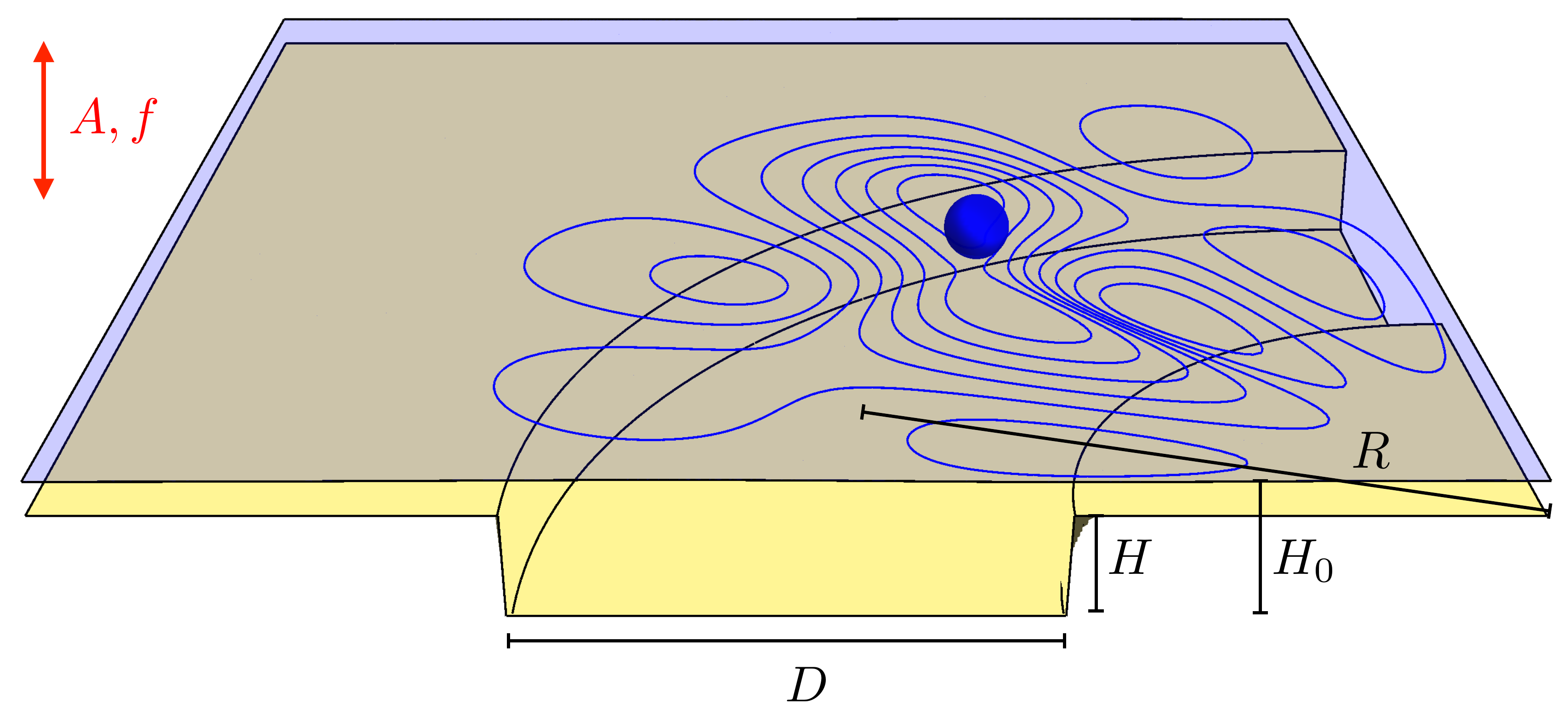}
\vskip -0.2 cm
\caption{{ \color{NavyBlue} \bf Sketch of the experiment.} A quarter of the small annular cavity of width $D$ and radius $R$. The oil level is adjusted to obtain a depth $H$ in the cavity and a thin layer $H_0$ elsewhere. A walking droplet tends to remain in the cavity. Contours of the liquid surface $\zeta(\vec r, t)$ are shown for illustrating that the propagation of waves mostly takes place in the cavity while evanescent waves are observed outside the cavity. }
\label{sketch}
\end{center}
\end{figure}

Since the occurrence of the Faraday instability is highly dependent of the liquid depth in the shallow regime, to design cavities is indeed a straightforward way to control the path of a walking droplet. We consider circular rings where droplets walk forever without reaching a boundary. The sketch of an annular cavity is given in Figure \ref{sketch}. The liquid depth in the center of the cavity is $H = 4 \, {\rm mm}$ and the width of the annular channel is $D = 7.5\, {\rm mm}$. Elsewhere, the liquid depth $H_0$ is fixed to 1 mm limiting the propagation of waves. At this depth, a droplet may bounce but cannot walk. Three ring cavities of different radii have been considered. 

Observations show that typical trajectories of single walkers are circles, but one should remind that at each bounce, the droplets take off and follow a tiny parabolic flight. There is no central force behind this circular trajectory. The fact that the droplet follow the ring is that the Faraday waves adopt the symmetry of the cavity. Along the ring, i.e. in the azimutal direction, waves can be approximated to sinusoidal standing waves, as expected for a 1d system. In the transverse direction, i.e. in the radial direction, the waves present antinodes in the center of the cavity and evanescent waves are strongly damped outside the ring. The droplet  remains in the central part of the cavity. Moreover, the speed $v_1$ of a single walker is fixed by the forcing parameters of the experiment and the geometry of the cavity. We checked that the speed $v_1 \approx 10 \, {\rm mm/s}$ of the walker is independent of the ring radius $R$. We will use this typical speed as a reference in the following.

\begin{figure}[h]
\vskip 0.1 cm
\includegraphics[width=8.5cm]{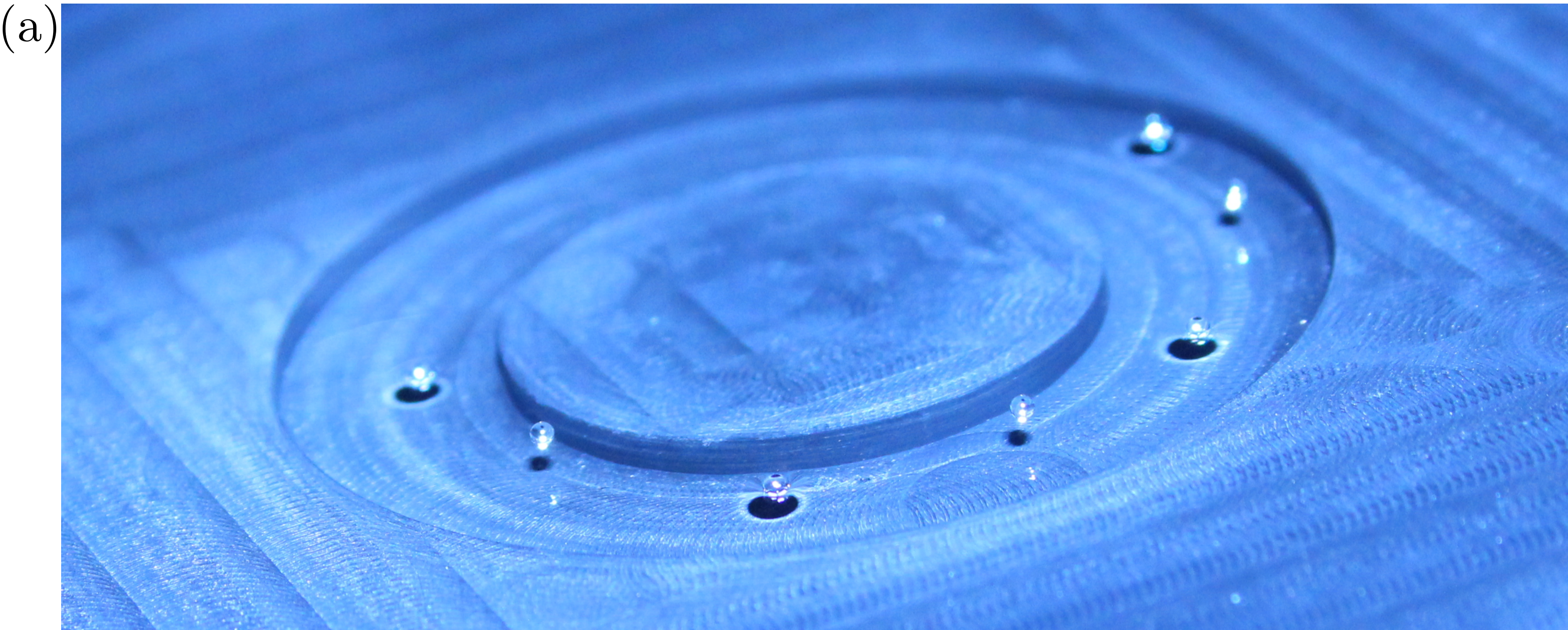}
\vskip 0.1 cm
\includegraphics[width=7cm]{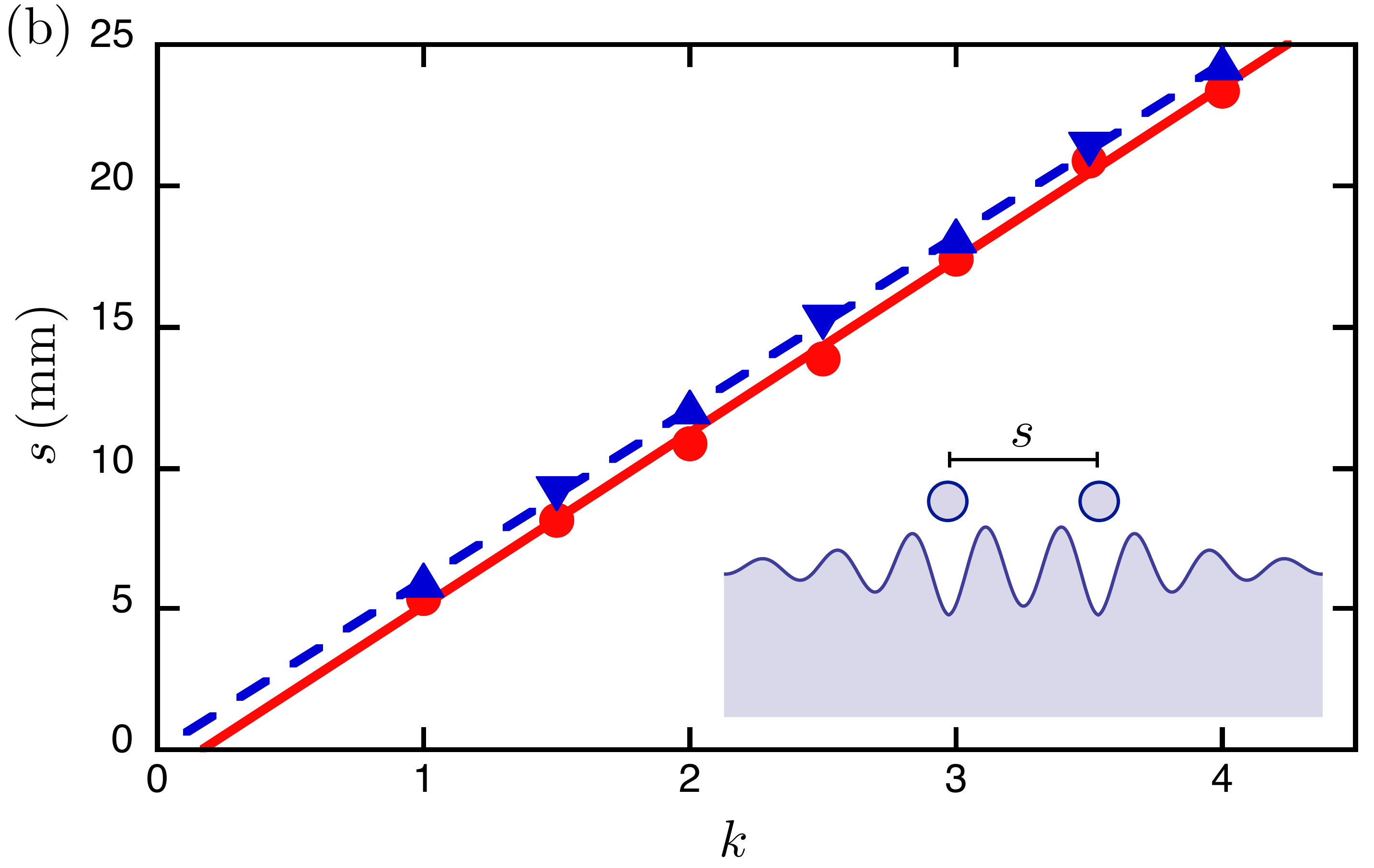}
\vskip -0.2 cm
\caption{{ \color{NavyBlue} \bf String of walking droplets.} (a) Picture of a group of 7 droplets in the small annular cavity. One may observe quantified interdistances and antisynchroneous bounces for the successive droplets. (b) Red dots represent the droplet interdistances as a function of the label $n$ describing the interaction mode. The line is a fit using Eq.(\ref{eq:lambdaF}). Error bars are not indicated since they are smaller than the symbol size.  Blue triangles are droplet interdistances given by the model for synchroneous (triangle up) or antisynchroneous bounces (triangle down). The dashed line is a fit using Eq.(\ref{eq:lambdaF}). }
\label{pdf}
\end{figure}

When two or more droplets are placed in the ring, they walk clockwise or counterclockwise. After a while, they start to interact. The result after long times is the formation of a string of walking droplets moving cooperatively along the ring. The most striking results is probably that in this final state they share a common and coherent wave. Indeed, the distances between droplets corresponds to multiple of $\lambda_F/2$ and bounces are (anti)synchronized. Figure \ref{pdf}(a) presents a picture of such a group made of $N=7$ droplets. In that string, two successive droplets are antisynchronized meaning that when the first one bounces, the second one is in a free flight, and conversely. On the picture, the shadows below each droplet illustrate this particular ordering. Strings of synchroneous droplets can be also created but their interdistances should be different, as seen below. 

Figure \ref{pdf}(b) shows the interdistances between successive droplets in various groups of 2 droplets that are launched randomly in an annular cavity. After a while, they form a group and one observes clearly specific distances being given by
\begin{equation}
s=(k-\epsilon_0)\lambda_F
\label{eq:lambdaF}
\end{equation}
that Couder and coworkers have already described for 2d systems \citep{Protiere2008,Borghesi2014}. The label $k$ is an integer or a half integer. The shift $\epsilon_0$ results from subtle interactions between droplets. The Figure \ref{pdf} shows a fit using Eq.(\ref{eq:lambdaF}) in order to determine $\lambda_F$ and $\epsilon_0$ which corresponds to values : $\lambda_F = 6.1 \pm 0.1 \, {\rm mm}$ and $\epsilon_0 = 0.18 \pm 0.02$. The first situation corresponds to $n=1$ for which the bounces are synchronized. Half integer values of $k$ correspond to antisynchronized values. The case $k=1/2$ is particular since the droplets start to orbit around each other and they follow complex trajectories. They finally coalesce or vanish in the vibrating bath. 

The coherent character of the wave propelling the string of walkers has been tested. When a string of a few droplets is formed, we used a needle to destroy one of the central droplets. The system appears to be unaffected by the disparition of this droplet and behaves exactly as before. The distances remain unchanged and the synchronicity is kept. Movies of that experiments are given in the supplementary materials, illustrating coherence effects. 

When the number $N$ of droplets increases, the coherent wave extension increases accordingly. At some point, the wave starts to self-interfere such that the system destabilizes. We checked that it is nearly impossible to form a complete ring of droplets moving cooperatively due to this effect. The only stable groups of droplets are found up to 12 droplets in the small ring. In the same spirit, modifying the acceleration will inevitably change the stability of the system. At high memory, i.e. when $\Gamma$ is close to $\Gamma_F$, the wave extension is more important leading also to interference and the possibility to destabilize the system. We avoid the possible collapse of the structure by keeping the system in a short memory regime.

\begin{figure}[h]
\begin{center}
\vskip 0.1 cm
\includegraphics[width=7cm]{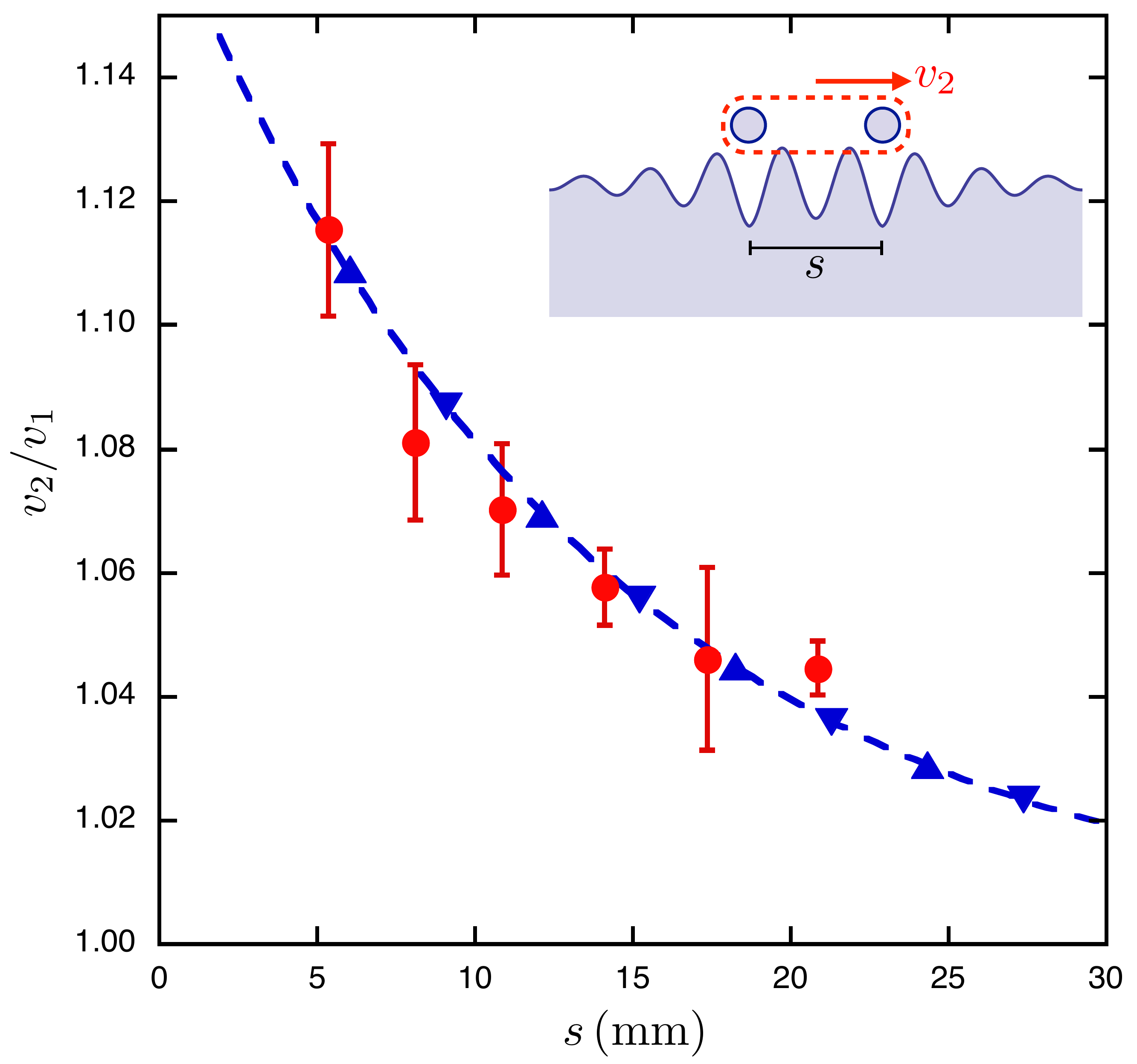}
\vskip -0.2 cm
\caption{{ \color{NavyBlue} \bf Speed for a pair of walkers.} (Dots) Speed of a droplet pair $v_2$ as a function of the distance $s$ between droplets. The speed is normalized by the speed $v_1$ of a single droplet. Error bars are indicated. (Triangles) The model explained in the main text returns quantified interdistances $s$ between droplets as well as specific speeds $v_2$ for both synchronous (triangle up) and antisynchronous (triangle down) cases. An excellent agreement is found between the model and the experimental data. The dashed curve is an exponential decay fitting the results from the model. }
\label{speeds2}
\end{center}
\end{figure}

The collective motion of the string was an unexpected result because in 2d only a few cases lead to a motion of the center of mass. It was recently proposed that bounded states exhibit speeds lower than the individual droplets freely moving along the surface \citep{Borghesi2014}. We will demonstrate herein a clear violation of this conjecture. Let us analyze the speed of the walkers resulting from the collective effects. Figure \ref{speeds2} shows the speed $v_2$ of the pair of bouncing droplets as a function of the distance $s$ separating them. The group speed is normalized by the speed $v_1$ of single droplets. The single droplet case should represent the asymptotic case when the interdistances are large enough to neglect interactions. Unexpectedly, the speeds $v_2$ obtained for droplet pairs are larger than $v_1$ in both synchronous and antisynchronous cases. The speed ratio is also seen to decrease as a function of $s$. The common coherent wave shared by droplets has therefore a large driving force as we will explain below. 

\begin{figure}[h]
\begin{center}
\vskip 0.1 cm
\includegraphics[width=7cm]{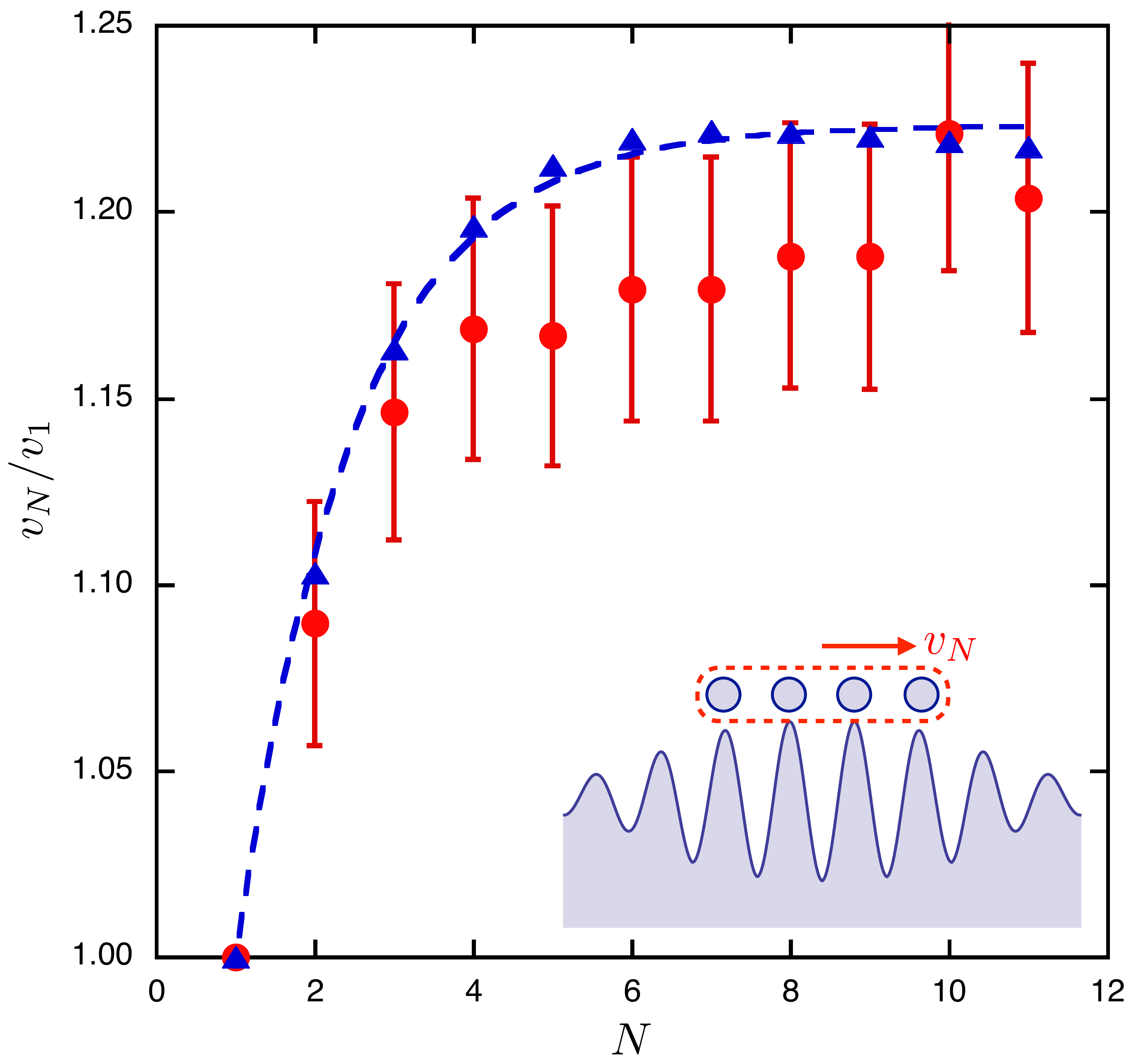}
\vskip -0.2 cm
\caption{{ \color{NavyBlue} \bf Speed for a string of walkers.} (Dots) Speed $v_N$ of a droplet string as a function of its number $N$ of components. The speed is normalized by the speed $v_1$ of single droplets. Error bars are indicated. The speed seems to saturate for large systems. (Triangles) The model described in the main text captures this effect for both synchronous (triangle up) and antisynchronous (triangle down) cases. An excellent agreement is found between the model and the experimental data.  The dashed curve is a guide for the eye. }
\label{speedsN}
\end{center}
\end{figure}

In order to emphasize the collective effects induced by wider coherent waves, we created groups of $N$ droplets being separated by a single wavelength $\lambda_F$. All droplets are therefore bouncing in a synchronized way. The normalized speed $v_N/v_1$ of the string increases with $N$, as shown in Figure \ref{speedsN}. The speed of the string seems to saturate for high $N$ values, meaning that there is a limit for coherence. The origin of this limit is due to the damping of Faraday waves. Indeed, above a droplet number (around $N=6$ in our experimental conditions) any additional droplet joining the string will not affect the global dynamics. We also created strings with a larger separation between successive walkers, like $3\lambda_F/2$. Similar results (not shown herein) are obtained but the saturation speed is lower than the previous case. 

The spontaneous formation of a string of walkers and its subsequent dynamics are captured by a model. We first assume that the standing waves associated to the successive droplet impacts are given by damped sine waves along the $s$-coordinate. We reduce the system to a 1d liquid profile $\zeta(s,t)$ along the ring. The driving horizontal force is supposed to be proportional to the slope of the liquid profile at each impact. Since the successive bounces are periodic events, separated by a time $\tau_F = 2/f$, a phenomenological strobed equation of motion is considered for each droplet $i$. The speed change $u_{n+1}^i-u_n^i $ of droplet $i$ between the $n$th and $(n+1)$th impacts is given by
\begin{equation}
u_{n+1}^i - u_n^i = - \gamma u_n^i - C_0 \left. {\partial \zeta^{ii} \over \partial s} \right|_{n+1} - C_1  \sum_{j \ne i}\left. {\partial \zeta^{ij} \over \partial s} \right|_{n+1},
\label{eq:motion}
\end{equation}
The first term represents some dissipation at bouncing with a parameter $\gamma$. The second term with a coefficient $C_0$ represents the interaction of the droplet $i$ at the $(n+1)$th impact with the waves produced by the same droplet at previous impacts. The other terms with a coefficient $C_1$ represents the interactions between the droplets in a string. Indeed, the droplet $i$ is also influenced by the waves created by the previous impacts of other droplets $j \ne i$. The values of the coefficients $C_0$ and $C_1$ could be different. Since we only focus on droplet motions at short memory regimes, the relevant information for the strobed equation (\ref{eq:motion}) is given by
\begin{equation}
\zeta^{ij} = \zeta_0 \cos \left({2 \pi (s_{n+1}^i-s_{n}^j) \over \lambda_F} \right) \exp \left(-{s_{n+1}^i-s_n^j \over \delta}\right)
\end{equation} where $\zeta_0$ is an arbitrary parameter. More details about the model are given in the supplementary information. 

For $N=2$ droplets, one obtains a set of two coupled equations similar to Eq.(\ref{eq:motion}). By considering that the speed of both droplets are identical, i.e. $u_n^1=u_n^2=v_2$, only one solution is found with quantified distances depending on either synchronous or antisynchronous bounces. The results are shown in Figure \ref{pdf}(b). The agreement with the experimental data is excellent, except for $\epsilon_0$ which remains close to zero in the model. Moreover, it can be shown that the speed of a group is larger than individual speeds. The group speed $v_2$ is seen to decay with the interdistance between two droplets. This is illustrated on Figure \ref{speeds2}. Extra speed for a pair of droplets is due to the constructive interference of waves emitted from both droplets. The parameters of the model are fixed to $\delta=2.1\lambda_F$, $C_0/\gamma \approx 0.03$ and $C_0 \approx 20 C_1$. Since the wave damping is already taken into account through $\delta$, the fact that $C_1$ is much lower than $C_0$ has an origin related to the non-trivial dissipation of waves in the cavity. Indeed, the self-interaction of a droplet with its wave is a local phenomenon while cross interactions between droplets involves propagation and reflection of waves along the cavity. Although different interaction coefficients should be considered, the coherent wave dynamics is well captured by our model. 

By generalizing the calculations to $N$ droplets, it is possible to estimate numerically the speed increase of $v_N$ reported in Figure \ref{speedsN}. Numerical results are shown on this plot (see triangles). Only qualitative agreement with the experimental data are obtained. Moreover, using different quantized distances for successive droplets, we are able to show that the asymptotic speed limit decreases with the distance between adjacent droplets. In fact, decoherence appears for distant droplets. This is also due to the low memory regime assumed herein.  

In summary, we have investigated bouncing droplets in a geometry confining them into a nearly 1d system. The wave emitted by a walker becomes a damped sine wave. We evidenced a remarkable feature for 1d walkers : they form group bouncing collectively leading to a faster motion than independent elements. Walkers have fascinated scientists for their peculiar behavior reminiscent of quantum like effects. The fact that coherent waves can be created push the analogy further. Moreover, 1d systems as considered herein are promising since droplets can be conveyed to collisions and other phenomena which are reminiscent of quantum information or wavefields.

\vskip 0.2 cm
{\bf Acknowledgments} -- This work was financially supported by the Actions de Recherches Concert\'ees (ARC) of the Belgium Wallonia- Brussels Federation under Contract No. 12-17/02.

\vskip 0.2 cm
{\bf Methods} -- The experimental conditions for the droplet and bath are the following. Identical tiny droplets are created by an automatic generator as fully described in \citep{Terwagne2013}. The resulting diameter of each droplet is close to 800 $\rm \mu m$. The liquid is silicon oil with a kinematic viscosity of 20 cSt, density $\rho=949 \, {\rm kg \, m^{-3}}$ and surface tension $\sigma = 20.6 \, {\rm mN \, m^{-1}}$. The container is shaken vertically by an electromagnetic system with a tunable amplitude $A$ and a fixed frequency $f=70 \, {\rm Hz}$. An accelerometer is fixed on the vibrating plate and delivers a tension proportional to the acceleration. The dimensionless maximum acceleration $\Gamma$ is tuned for finding the bouncing and walking regimes close to the Faraday instability. The acceleration is always kept at $\Gamma = 0.95 \Gamma_F$ in our experiments. This corresponds to a so-called ``low memory regime" since Faraday waves are damped such that the liquid surface $\zeta(\vec r, t)$ keeps the trace of a few previous impacts. The wavelength has been estimated to about $\lambda_F \approx 6 \, {\rm mm}$.

First, we reduce the walker trajectories to a nearly 1d system. Since the occurrence of the Faraday instability is highly dependent of the liquid depth in the shallow regime, to design cavities is indeed a straightforward way to control the path of a walking droplet. We consider circular rings where droplets walk forever without reaching a boundary. The sketch of an annular cavity is given in Figure \ref{sketch}. The liquid depth in the center of the cavity is $H = 4 \, {\rm mm}$ and the width of the annular channel is $D = 7.5\, {\rm mm}$, i.e. the width is in between $\lambda_F$ and $\frac{3}{2}\lambda_F$. The optimal width $D$ and depth $H$ of such a cavity was determined in an independent work. Elsewhere, the liquid depth $H_0$ is fixed to 1 mm limiting the propagation of waves. At this depth, a droplet may bounce but cannot walk. Each ring cavity is characterized by an inner radius $R_{in}$ and an outer radius $R_{out}=R_{in}+D$. Three different radii have been considered : (small ring) $R_{in}=10 \, {\rm mm}$, (medium ring) $R_{in}=37.5 \, {\rm mm}$ and (large ring) $R_{in}=65 \, {\rm mm}$ in our experiments. A Pixelink camera records walker trajectories and a Python code tracks drop motion in real time. 

\vskip 0.2 cm
{\bf Supplementary Information}

\begin{enumerate}

\item A movie is provided. It illustrates all the effects reported herein : speed of droplets independent of the ring size, string formation when a few walkers are placed in an annular cavity and the speed increase seen when droplets are in interaction. 

\item The model for droplet walking and interacting in the cavity is detailed in a separate file. 
\end{enumerate}

\vskip 0.2 cm
{\bf Authors Contributions Statement} -- BF collected and analyzed experimental data. Physical interpretations were provided by MH and NV. This manuscript was written by BF and NV. 

\vskip 0.2 cm
{\bf Competing Interests} -- The authors declare that they have no competing financial interests.


\end{document}